# The development of ferromagnetism in the doped topological insulator $Bi_{2-x}Mn_xTe_3$

Y. S. Hor[1], P. Roushan[2], H. Beidenkopf[2], J. Seo[2], D. Qu[2], J. G. Checkelsky[2], L. A. Wray[2], Y. Xia[2], S.-Y. Xu[2], D. Qian[2], M. Z. Hasan[2], N. P. Ong[2], A. Yazdani[2], and R. J. Cava[1]

[1]Department of Chemistry, Princeton University, Princeton, New Jersey, USA.

[2]Department of Physics, Princeton University, Princeton, New Jersey, USA.

**Abstract**

The development of ferromagnetism in Mn-doped $Bi_2Te_3$ is characterized through measurements on a series of single crystals with different Mn content. Scanning tunneling microscopy analysis shows that the Mn substitutes on the Bi sites, forming compounds of the type $Bi_{2-x}Mn_xTe_3$, and that the Mn substitutions are randomly distributed, not clustered. Mn doping first gives rise to local magnetic moments with Curie-like behavior, but by the compositions $Bi_{1.96}Mn_{0.04}Te_3$ and $Bi_{1.91}Mn_{0.09}Te_3$ a second order ferromagnetic transition is observed, with $T_C \sim$ 9-12 K. The easy axis of magnetization in the ferromagnetic phase is perpendicular to the $Bi_2Te_3$ basal plane. Thermoelectric power and Hall effect measurements show that the Mn-doped $Bi_2Te_3$ crystals are p-type. Angle resolved photoemission spectroscopy measurements show that the topological surface states that are present in pristine $Bi_2Te_3$ are also present in ferromagnetic Mn-doped $Bi_{2-x}Mn_xTe_3$, and that the dispersion relations of the surface states are changed in a subtle fashion.

## Introduction

Bi$_2$Te$_3$ has long been studied as the parent compound of a family of excellent ambient temperature thermoelectric materials (see e.g. refs 1,2). Recently, however, Bi$_2$Se$_3$ and Bi$_2$Te$_3$ have been of renewed interest because they are bulk topological insulators (see e.g. refs 3-7), hosting Dirac-like conducting surface states [8-10]. The doping of topological insulators to decrease bulk carrier concentration [11] and induce superconductivity [12] has been performed in order to investigate various aspects of the surface states in these materials, including their potential applicability in novel electronic devices. Of particular interest in future work will be the effects of magnetic impurities and ferromagnetism on the topological surface states. Here we report the development of ferromagnetism near 12 K as a function of Mn doping in single crystals of Bi$_{2-x}$Mn$_x$Te$_3$. Our work builds on a previous report of ferromagnetism in Bi$_{1.98}$Mn$_{0.02}$Te$_3$ [13], and other work on the dilute ferromagnetic V-VI semiconductors Bi$_{2-x}$Fe$_x$Te$_3$ [14] (a very weak ferromagnet) and Sb$_{2-x}$V$_x$Te$_3$ [15]. We show quantitatively how the Bi$_{2-x}$Mn$_x$Te$_3$ system develops ferromagnetism as a function of doping, characterize the material at the nanoscale by scanning tunneling microscopy (STM), and present preliminary angle resolved photoemission spectroscopy (ARPES)  characterization of the effects of doping on the surface states.

## Experimental

High purity elemental Bi (99.999 %), Mn (99.99 %), and Te (99.999 %) were used for the Bi$_{2-x}$Mn$_x$Te$_3$ crystal growth, employing nominal x values of 0.005, 0.01, 0.02, 0.05 and 0.1. Because Mn reacts with quartz, a two-step melting method, described in ref. 11, was necessary for successful modified Bridgeman crystal growth. The crystal growth for Bi$_{2-x}$Mn$_x$Te$_3$ involved cooling from 950 to 550 ºC over a period of 24 hours and then annealing at 550 ºC for 3 days;

silver-colored single crystals were obtained. The crystals were confirmed to be single phase and identified as having the rhombohedral $Bi_2Te_3$ crystal structure by X-ray diffraction using a Bruker D8 diffractometer with Cu Kα radiation and a graphite diffracted beam monochromator. X-ray diffraction patterns of the cleaved crystals oriented with basal plane normals bisecting the incident and diffracted beam directions showed only the (0 0 3), (0 0 6), and (0 0 15) peaks (hexagonal setting), indicating that the cleaved surface is oriented perpendicular to the hexagonal *c* axis. The *c*-axis parameter does not change substantially with Mn doping: we refine the values to be 30.488(2) Å for $Bi_2Te_3$ and 30.467(3) Å for the highest Mn-doped sample. Elemental analysis (Galbraith Laboratories) indicated that the Mn contents in the single crystals are very close to the nominal values, with the correspondences: nominal x = 0.005, 0.01, 0.02, 0.05 and 0.1 are equal to the true x = 0.005, 0.01, 0.02, 0.04, and 0.09 respectively. The analytically determined Mn concentrations are employed in this report. DC magnetization measurements were performed on a Quantum Design Physical Property Measurement System (PPMS). Temperature dependent resistivity measurements were carried out in the PPMS using the standard four-point probe technique with silver paste cured at room temperature used for the contacts. Seebeck coefficient measurements were conducted using a homemade probe with an MMR Technologies SB100 Seebeck measurement system. Hall Effect measurements were performed between 5 and 200 K in a home-built apparatus. In all cases, the electric- and thermal-currents were applied in the basal plane (hexagonal *ab* plane) of the crystals. The $Bi_{2-x}Mn_xTe_3$ crystal surfaces were characterized with a homemade cryogenic scanning tunneling microscope (STM) at 4.2 K under ultra-high vacuum conditions. Angle resolved photoemission spectroscopy (ARPES) data were measured on *in situ* cleaved crystal surfaces at 15 K, in a vacuum maintained below $6 \times 10^{-11}$ torr, at the Advanced Light Source, beam lines 10 and 12.

**Results and Discussion**

The temperature dependent magnetic susceptibilities, $\chi = M/H$, measured in an applied field of 1 kOe (measured magnetizations ($M$) are linearly dependent on applied magnetic field ($H$) up to 1 kOe) are shown in Figure 1 for the $Bi_{2-x}Mn_xTe_3$ crystals for x = 0, 0.005, 0.01, 0.02, 0.04 and 0.09, measured with $H$ perpendicular to the $c$ axis. The susceptibility is negative at high temperatures for x < 0.04, an indication of the weakness of the paramagnetic contribution from the Mn, which becomes significant only at low temperatures. There is no indication of a ferromagnetic transition in the susceptibility measurements for x < 0.04. The system becomes ferromagnetic when the Mn concentration is increased to x = 0.04, evidenced in these data by a relatively large susceptibility at low temperatures. The inset of Figure 1 emphasizes the low temperature susceptibilities for the series; for the crystals with x = 0.04 and 0.09 in $Bi_{2-x}Mn_xTe_3$, paramagnetic behavior is seen for the full temperature range between 300 K and $T_C$.

The susceptibility $\chi$ can be fit to the Curie-Weiss law, $\chi - \chi_0 = C/(T - \theta)$, where $\chi_0$ is the temperature independent term, $C$ is the Curie constant, and $\theta$ is the Weiss temperature. Figure 2 shows the low temperature inverse susceptibility plots, $1/(\chi - \chi_0)$ vs. T, for all samples. The inverse susceptibilities for x = 0.005, 0.01, and 0.02 show straight line behavior, as expected for the presence of weak local moments due to the Mn dopants. For the lowest x, a very small negative $\theta$ (-0.7 K) is observed, which becomes slightly more positive with increasing Mn content. In contrast, the data for crystals of composition x = 0.04 and 0.09 follow the Curie-Weiss law with positive $\theta$s of 11 and 13 K, respectively. Figure 3 shows the temperature dependence of the inverse susceptibility for the $Bi_{2-x}Mn_xTe_3$ crystals normalized on the basis of moles of Mn for x = 0.005, 0.01, 0.02, 0.04, and 0.09. From this data, the effective moment per Mn, $p_{eff}$ can be obtained from $p_{eff} = (7.99C)^{1/2}$. As the Mn doping initially increases, the effective moment per mol-Mn decreases, but for most of the range, x = 0.02 - 0.09, $p_{eff}$ per Mn does not change much, as seen in the inset of Figure 3, giving the magnitude of $p_{eff}$ ~ 4 $\mu_B$ per mol-Mn.

To better determine the temperatures of the ferromagnetic transitions, we apply the criterion employed for disordered systems [16, 17] and present Arrott plots in Figures 4 and 5 for $Bi_{1.96}Mn_{0.04}Te_3$ and $Bi_{1.91}Mn_{0.09}Te_3$. In such plots, $H/M$ versus $M^2$ should go as $H/M = a'(T - T_C) + b'M^2 + c'M^4$ at values of $H$ above those of the magnetic hysteresis. At $T_C$, the intercept changes sign. For second order transitions, $b'$ is positive. Before each run, the samples were heated above their $T_C$ and cooled to the measuring temperature under zero field to ensure perfect demagnetization. The data in Figure 4 indicate that the ferromagnetic transition in $Bi_{1.96}Mn_{0.04}Te_3$ is second order as expected, with $T_C$ determined as ~9 K. The Arrott plot for $Bi_{1.91}Mn_{0.09}Te_3$, Figure 5, similarly shows a second order ferromagnetic transition with a $T_C$ of approximately 12 K.

Figure 6 shows the field dependent magnetization curves at 1.8 K for a $Bi_{1.91}Mn_{0.09}Te_3$ crystal, where the applied field is parallel to the $c$ axis, $H//c$, and perpendicular to the $c$ axis, $H\perp c$. The magnetization for $H//c$ has a very steep initial slope, whereas the magnetization for $H\perp c$ increases much more slowly. This indicates that the $c$ axis is the easy axis of magnetization, with spontaneously ordered magnetic moments in the ferromagnetic phase aligned perpendicular to the $Bi_2Te_3$ basal plane. This moment direction is the same as is seen in V-doped $Sb_2Te_3$ [15]. A narrow hysteresis loop is seen for $H//c$, as shown in the inset of Figure 6. The small coercivity, $H_C$ ~ 350 Oe, indicates that the sample is a soft ferromagnet. $H_C$ for our crystal of $Bi_{1.91}Mn_{0.09}Te_3$ is much smaller than is seen for V-doped $Sb_2Te_3$, where $H_C$ is ~ 12 kOe at 2 K [15]. The saturated magnetic moment, reached by about 12 kOe for $H//c$ and 20 kOe for $H\perp c$ at 1.8 K, is 1.5 $\mu_B$ per mol Mn.

The temperature dependence of the basal plane resistivities for $Bi_{2-x}Mn_xTe_3$ crystals for x = 0.005, 0.02, and 0.09 are shown in Figure 7. As Mn-doping increases, the resistivity of $Bi_{2-x}Mn_xTe_3$ increases, though the metallic-like behavior typically observed in heavily doped small band gap semiconductors is maintained. The increased resistivity at low temperatures in the more heavily doped samples is due to the presence of a larger residual resistivity, indicative of strong carrier scattering from the doped Mn. There is a sharp drop in resistivity for $Bi_{1.91}Mn_{0.09}Te_3$ at ~13 K that accompanies the ferromagnetic transition seen in the susceptibility measurements. The resistivity anomaly is pushed to higher temperature and substantially broadened under a 10 KOe magnetic field applied parallel to the *c*-axis, shown in the inset of Figure 7. Broad drops in the resistivities for x = 0.02 and 0.09 crystals are observed at temperatures around 250 K. The origin of these resistivity drops is unknown; further investigation is suggested. The Hall resistivity $\rho_{yx}$ measured on crystals with x = 0.04 and 0.09 shows that the carriers are hole-like with Hall density $p = 1/e\rho_{yx} = 5 \times 10^{18}$ and $7 \times 10^{19}$ cm$^{-3}$, respectively (Figure 8). In the x = 0.04 sample, weak quantum oscillations resolved in the derivative $d\rho_{yx}/dH$ can be observed as depicted in Figure 9. The period of the oscillations, $\Delta(1/H) = 0.046$ T$^{-1}$, corresponds to a small FS pocket with Fermi wavevector $k_F = 0.047$ Å$^{-1}$ and carrier density $3.5 \times 10^{18}$ cm$^{-3}$, a value slightly smaller than the measured Hall density. Figure 10 shows the thermoelectric power as a function of temperature for the $Bi_{2-x}Mn_xTe_3$ crystals. $Bi_{2-x}Mn_xTe_3$ is p-type in the whole range of temperatures for Mn substitutions up to x = 0.09. The data in Figure 10 show that Mn doping on the Bi site dramatically reduces the Seebeck coefficient, from 280 to 80 μVK$^{-1}$ at room temperature. As temperature decreases, the Seebeck coefficients for Mn-doped $Bi_2Te_3$ monotonically decrease.

In order to directly visualize the effects of Mn doping in $Bi_2Te_3$ at the atomic scale, we have characterized the $Bi_{2-x}Mn_xTe_3$ (001) surface with a cryogenic STM. Due to the weak van-der-Waals bonding between the adjacent Te layers, the uppermost layer exposed after *in situ* cleavage is a triangular lattice of Te atoms. A typical STM topographic image of the surface of $Bi_{1.91}Mn_{0.09}Te_3$ is presented in Figure 11(a). The black triangles are identified as substitutional Mn. Figures 11(b) and (c) show zoomed-in views over a triangular local density of states (LDOS) suppression associated with Mn dopants, in which the position of Te atoms on the topmost layer are marked by blue (dark) circles. Below the surface Te layer, the Bi atoms form an interpenetrating triangular lattice (sites denoted by red (medium) circles) centered in the middle of the triangle formed by the Te atoms in the surface layer. The triangular suppression in the LDOS that extends over several lattice sites is in the same lattice position expected for a Bi atom; therefore, we conclude that Mn substitutes primarily for Bi upon doping. In the topographies of the filled states, as shown in Figure 11(c), Mn dopants are also centered on the Bi sites, forming a smaller region of suppression and a bright halo signifying enhanced LDOS. We attribute the formation of this bright region to local band bending due to negative charge that localizes about the Mn ion. This is as expected around substitutional donor atoms. In addition to the Mn dopants in the Bi layer, faint clover-like patterns are visible in topographic images. These could be substitutional Mn atoms in deeper Bi layers, but their accurate identification requires further study. The number of contrast features associated with Mn atoms seen in the top most Bi layer appears to be significantly less than that expected from the measured Mn concentration; the reason for this is not known.

The STM topographic images can be used to study the spatial distribution of the Mn dopant atoms. Given the high concentration of Mn in ferromagnetic samples of $Bi_{2-x}Mn_xTe_3$, the formation of Mn clusters, if present, could indicate a completely different mechanism for the observed ferromagnetism. By locating the position of substitutional Mn, STM provides a direct visual method for examining the tendency to form clusters from the atomic to sub-micron scale. As discussed above, the substitutional Mn atoms in the closest Bi layer to the surface appear as triangular suppression of the LDOS in topographic images. After finding the position of the individual Mn dopants in one of such Bi layers, the correlation between Mn pairs was calculated for large fields of view (>1000 Å). The plot presented in Figure 11(d) is the cumulative probability of random pair separation; i.e., for a given value of $r$, $P(<r)$ is the probability of two randomly chosen Mn dopants being less than a distance $r$ from one another. For homogeneously distributed dopants, this probability scales with the area ($r^2$), and the dopants fill the field of view, leaving no void area. The measured probability shows a power law behavior with a power very close to 2 for an extended range of distances ($10 Å < r < 200 Å$), indicating the absence of clustering in $Bi_{2-x}Mn_xTe_3$ for concentrations as high as x = 0.09. In contrast to the single exponent observed here, the tendency toward cluster formation would result in different exponents at different length scales, starting with values higher than 2 at short distances and smaller than 2 at large distances. Five topographies from randomly chosen areas of the surface several hundreds of microns apart were also analyzed and yielded similar results.

The electron acceptor nature of Mn dopants can be seen in the STM from the evolution of the LDOS with doping. The LDOS of the undoped sample, shown in Figure 12 (blue (darker) line), consists of bulk conduction and valence band contributions that decay towards the bulk gap. Within the gap, the topological surface states of $Bi_2Te_3$ are the sole contributors to the LDOS. Upon Mn doping, the LDOS (red (lighter) line) shifts as a whole to higher energies, signifying the reduced density of unbound electrons in the Mn doped sample. Accordingly the Fermi energy, $E_F$, shifts with doping toward the valence band, rendering the sample p-type. The surface states remain present in the gap at 4.2 K in the Mn doped crystal, well within the ferromagnetically ordered temperature regime.

Finally, we have studied the topological surface state band structure with angle resolved photoemission spectroscopy (ARPES) to further characterize changes in the electronic behavior of the system on Mn doping. The surface states in undoped $Bi_2Te_3$ are lightly electron doped, with approximately 0.001 charge carriers in the Brillioun zone, as estimated from the Fermi surface area (Figures 13(a and b)). The surface states remain present at the x = 0.09 Mn doping level (Figure 13(c)). At the temperature of the measurements, 15 K, the material is very close to its full ferromagnetic transition temperature, so short range ferromagnetic ordering is present and strong in the bulk phase. The Mn substitution results in slight hole doping of the surface states, decreasing the chemical potential by approximately 180 meV to beneath the surface state Dirac point, leaving fewer than 0.001 hole-like charge carriers per surface Brillouin zone. The ARPES data show that the surface state characteristics are altered by the doping; a detailed description of doping-induced changes in the surface state dispersion will be reported elsewhere.

**Conclusions**

The development of ferromagnetism at low temperatures in Mn-doped $Bi_2Te_3$ has been followed as a function of dopant concentration. Ferromagnetism sets in at 2 % Mn substitution for Bi, x = 0.04 in $Bi_{2-x}Mn_xTe_3$, with a maximum $T_C$ of about 12 K seen for $Bi_{1.91}Mn_{0.09}Te_3$. The ordered ferromagnetic moments are aligned with the easy axis perpendicular to the $Bi_2Te_3$ basal plane. The absence of Mn clusters indicates that the system as a true dilute ferromagnetic semiconductor. The moment per $Bi_{2-x}Mn_xTe_3$ formula unit is substantially larger than is observed in the analogous Fe-doped $Bi_2Te_3$ system [14]. Important to the study of topological surface states, we have shown that for ferromagnetic samples the surface states remain present, though altered in subtle ways. It will be of substantial future interest to characterize the impact of the magnetism on the topological surface states in this class of compounds through further measurements.

**Acknowledgments**

This research was supported by the NSF MRSEC program, grant DMR-0819860. P. R. and D. Q. acknowledge NSF graduate fellowships. We thank M.M. Bandi for fruitful discussions.

**Figure Captions**

**Figure 1** (color online). Zero-field cooled (ZFC) temperature-dependent DC magnetic susceptibility $\chi$ measured at 1 kOe applied magnetic field for the $Bi_{2-x}Mn_xTe_3$ crystals. The magnetic susceptibility in the region of 0 to 30 K is shown in the inset.

**Figure 2** (color online). Temperature dependence of the inverse susceptibility for x = 0.005, 0.01, 0.02, 0.04, and 0.09 of $Bi_{2-x}Mn_xTe_3$.

**Figure 3** (color online). Temperature dependence of the inverse susceptibility per mole of Mn for x = 0.005, 0.01, 0.02, 0.04, and 0.09 of $Bi_{2-x}Mn_xTe_3$. The effective magnetic moment per mole of Mn is shown in the inset.

**Figure 4** (color online). Arrott plots at various temperatures near $T_C$ for $Bi_{1.96}Mn_{0.04}Te_3$. The ferromagnetic transition is of second order at $T_C \sim 9$ K.

**Figure 5** (color online). Arrott plots for $Bi_{1.91}Mn_{0.09}Te_3$ show the second order ferromagnetic transition with a $T_C$ of about 12 K.

**Figure 6** (color online). Magnetic field dependent magnetization of a $Bi_{1.91}Mn_{0.09}Te_3$ crystal under the applied field parallel to the $c$ axis, $H//c$ and perpendicular to the $c$ axis, $H \perp c$ at $T = 1.8$ K. The inset shows the hysteresis $MH$ loops of this $Bi_{1.91}Mn_{0.09}Te_3$ crystal for both field directions.

**Figure 7** (color online). Temperature-dependent resistivities for x = 0.005, 0.02, and 0.09 of $Bi_{2-x}Mn_xTe_3$. The inset shows the resistivity of $Bi_{1.91}Mn_{0.09}Te_3$ crystal in 0 and 10 kOe applied magnetic fields.

**Figure 8** (color online) The Hall resistivity $\rho_{yx}$ vs. field $H$ in $Bi_{2-x}Mn_xTe_3$ with x = 0.04 (upper panel) and x = 0.09 (lower panel). $\rho_{yx}$ is hole-like, and displays a slight change of slope in the weak-field limit. The hole density $p = H/e\rho_{yx}$ inferred at 8 T and 20 K equals $5 \times 10^{18}$ cm$^{-3}$ and $7 \times 10^{19}$ cm$^{-3}$, for x = 0.04 and 0.09, respectively.

**Figure 9** (color online). Quantum oscillations extracted from the derivative of the Hall resistivity $d\rho_{yx}/dH$ measured at 5 K in Bi$_{2-x}$Mn$_x$Te$_3$ with x = 0.04. The period of the oscillations $\Delta(1/H) = 0.046$ T$^{-1}$ gives a (spherical) FS volume of wavevector $k_F = 0.047$ Å$^{-1}$, corresponding to a carrier density, $3.5 \times 10^{18}$ cm$^{-3}$, slightly smaller than $p = H/e\rho_{yx}$.

**Figure 10** (color online). Temperature dependent Seebeck coefficients for Bi$_{2-x}$Mn$_x$Te$_3$ crystals.

**Figure 11** (color online). STM topographic image of Bi$_{1.91}$Mn$_{0.09}$Te$_3$, identifying the atomic substitutional site of Mn dopants, and the absences of Mn clustering. (a) STM topograph (+250 meV, 40 pA) of the Bi$_{1.91}$Mn$_{0.09}$Te$_3$ (001) surface over a 1000 Å by 1000 Å area. Substitutional Mn atoms appear as triangular suppressions of the LDOS. (b) and (c) Zoom-in topographies over Mn dopants in an area of 30 Å by 30 Å for unoccupied (+500 mV, 30 pA) and filled states (-500mV, 30 pA). The position of top Te layer, Bi layer, and substitutional Mn are shown by blue, magenta, and green circles on the topographies. (d) Study of the possibility of clustering by calculating the probability of correlation between Mn pairs in different locations. For a randomly chosen pair of Mn, $P(<r)$ gives the probability of the pair having a separation less than $r$. The calculated probability scales very close to $r^2$ (solid line) for an extended range of distances, demonstrating the uniform distribution of Mn dopants.

**Figure 12** (color online). Comparison of the local density of states (LDOS) for pristine Bi$_2$Te$_3$ (blue (darker) line) and Bi$_{1.96}$Mn$_{0.04}$Te$_3$ (red (lighter) line) as measured by STM at 4.2 K. Valence band (V.B.) and conduction band (C.B.) states are seen at the low and high bias values, and surface states (S.S.) are seen in the band gap at intermediate bias conditions. The chemical potential is shifted by about 150 mV toward the valence band in the Mn-doped sample, indicating the p-type character of the Mn dopants. The surface states are present in Mn-doped $Bi_2Te_3$ below its bulk ferromagnetic $T_C$.

**Figure 13** (a) The Fermi surface of the surface states of undoped $Bi_2Te_3$ is shown, with a complete map of momentum space in the inset revealing only ~0.001 conducting electrons in the surface Brillouin zone. (b-c) Energy-resolved ARPES measurements show that the chemical potential of $Bi_{1.91}Mn_{0.09}Te_3$ ("$\mu_{Mn}$") is lowered by approximately 180 meV compared to undoped $Bi_2Te_3$ ("$\mu_0$"), resulting in some changes to the surface states, and a transition from electron-like to hole-like charge carriers (~0.001 holes/BZ) in the surface states. Data were taken approximately 30 minutes after cleavage.

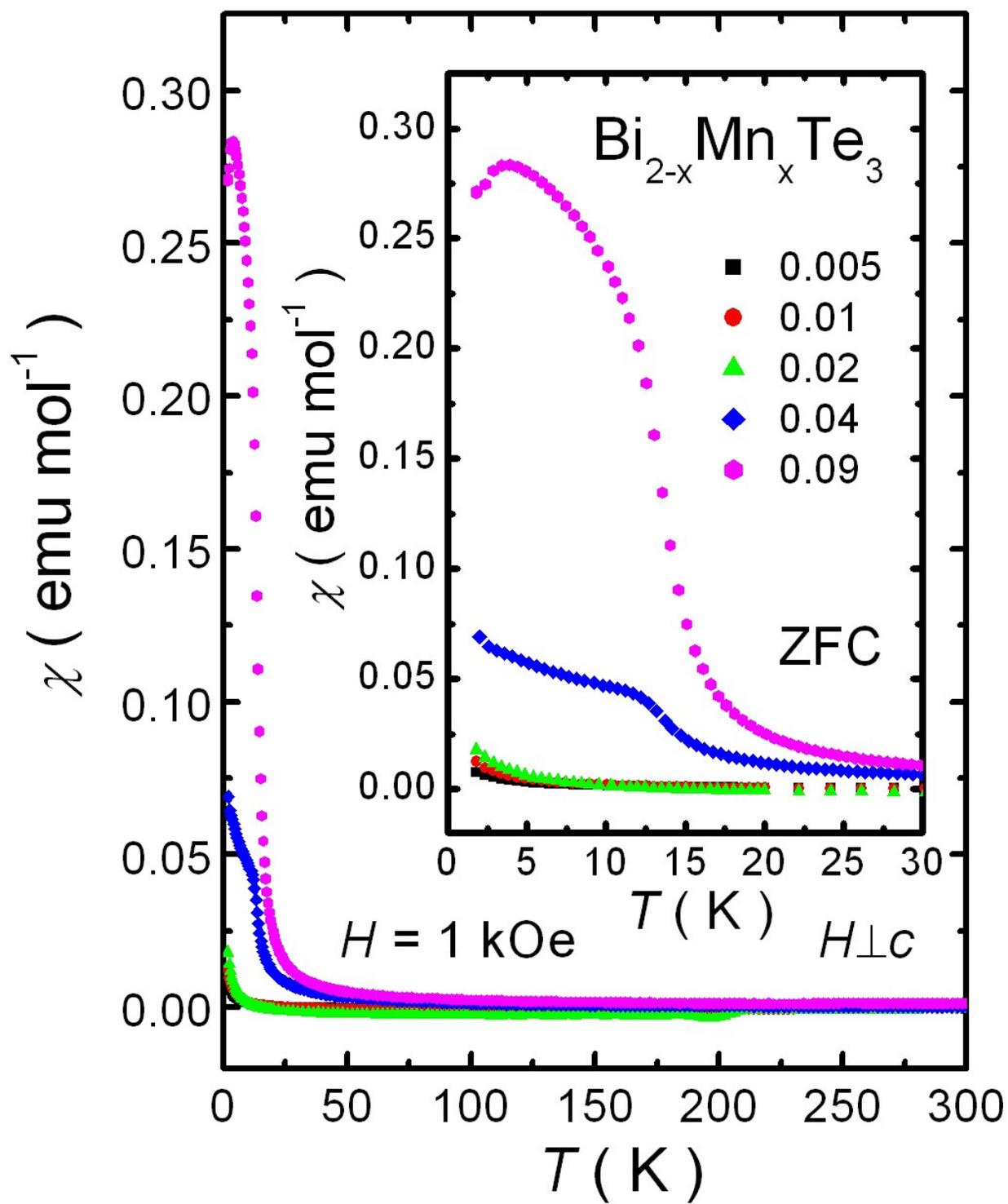

Figure 1

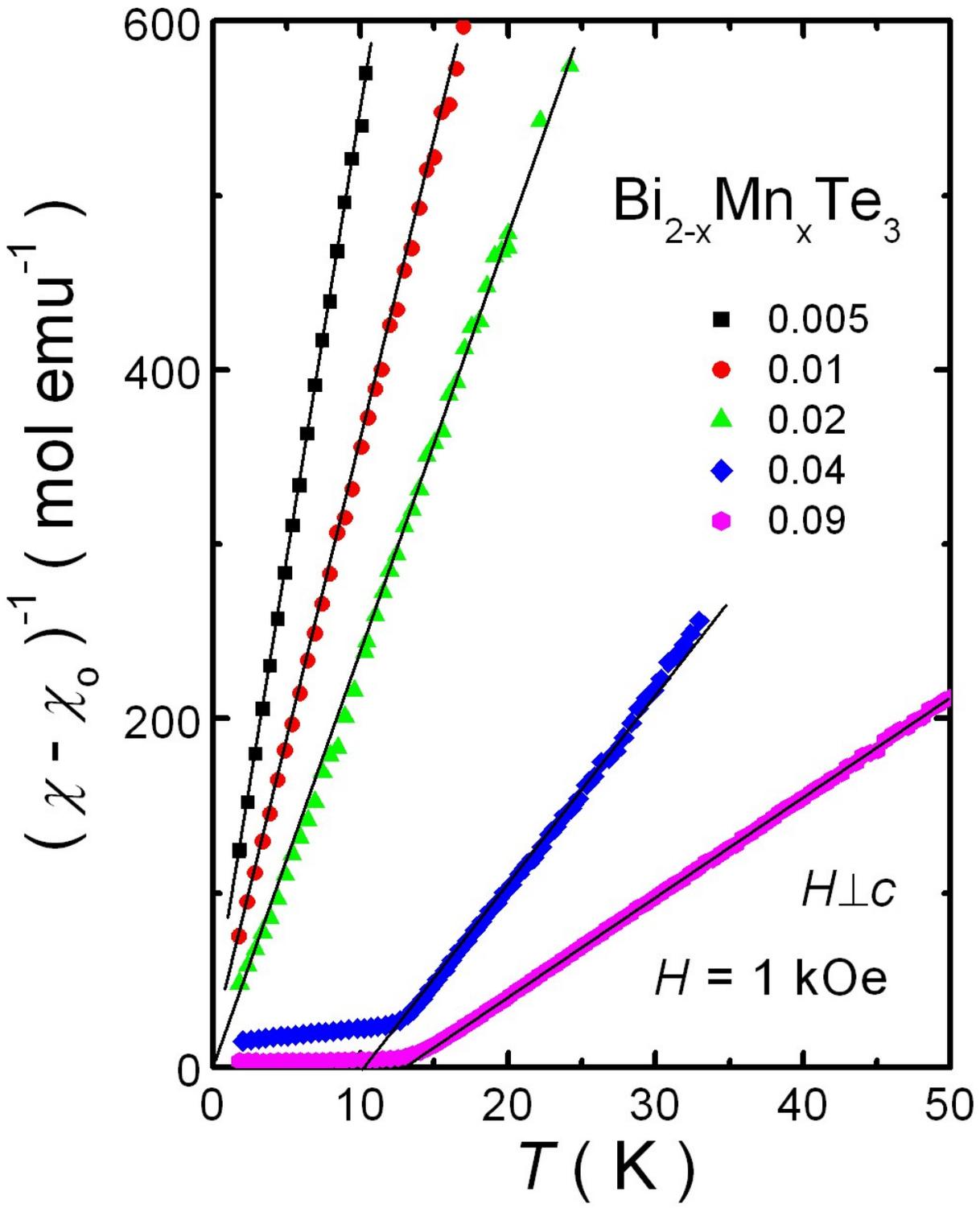

Figure 2

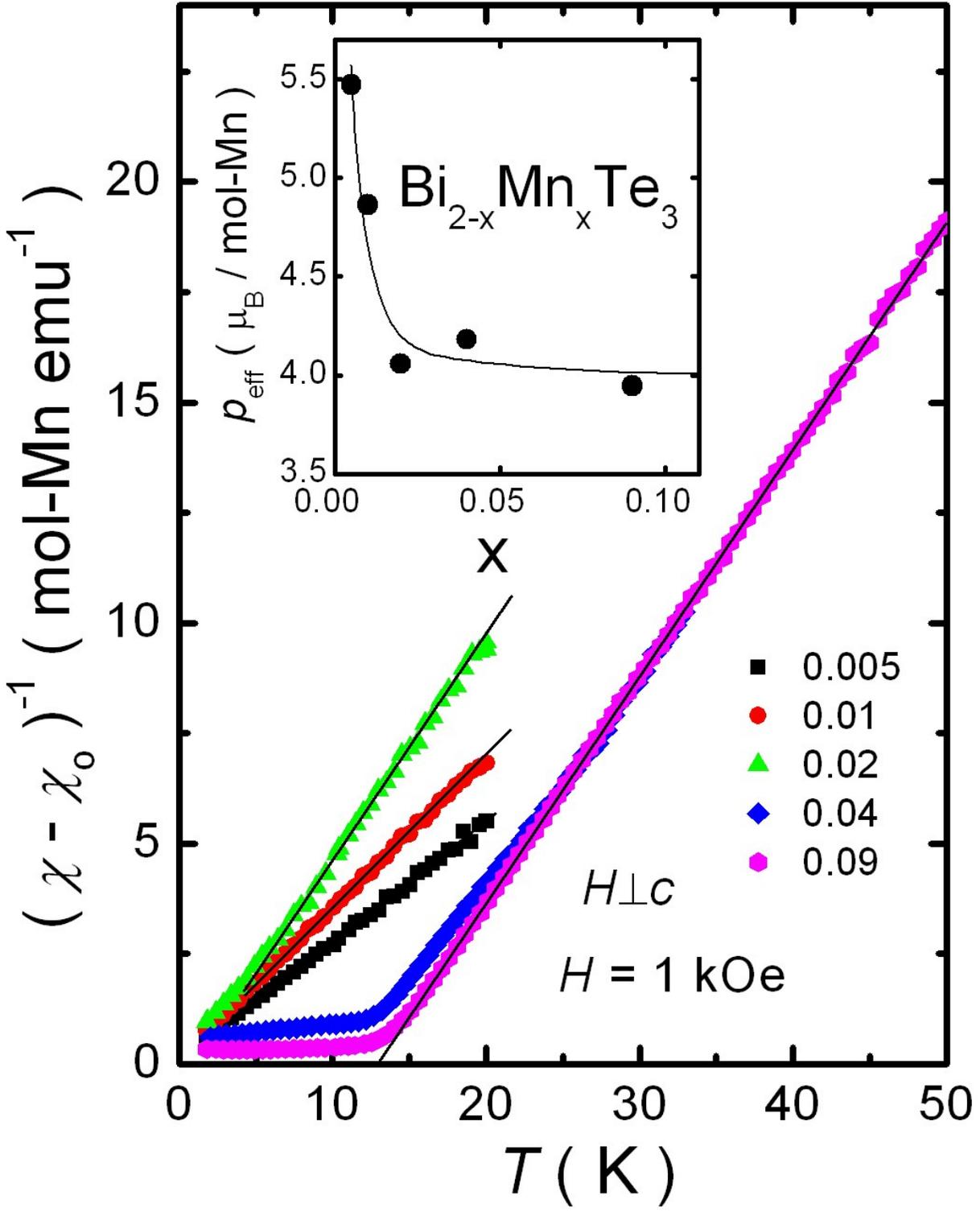

Figure 3

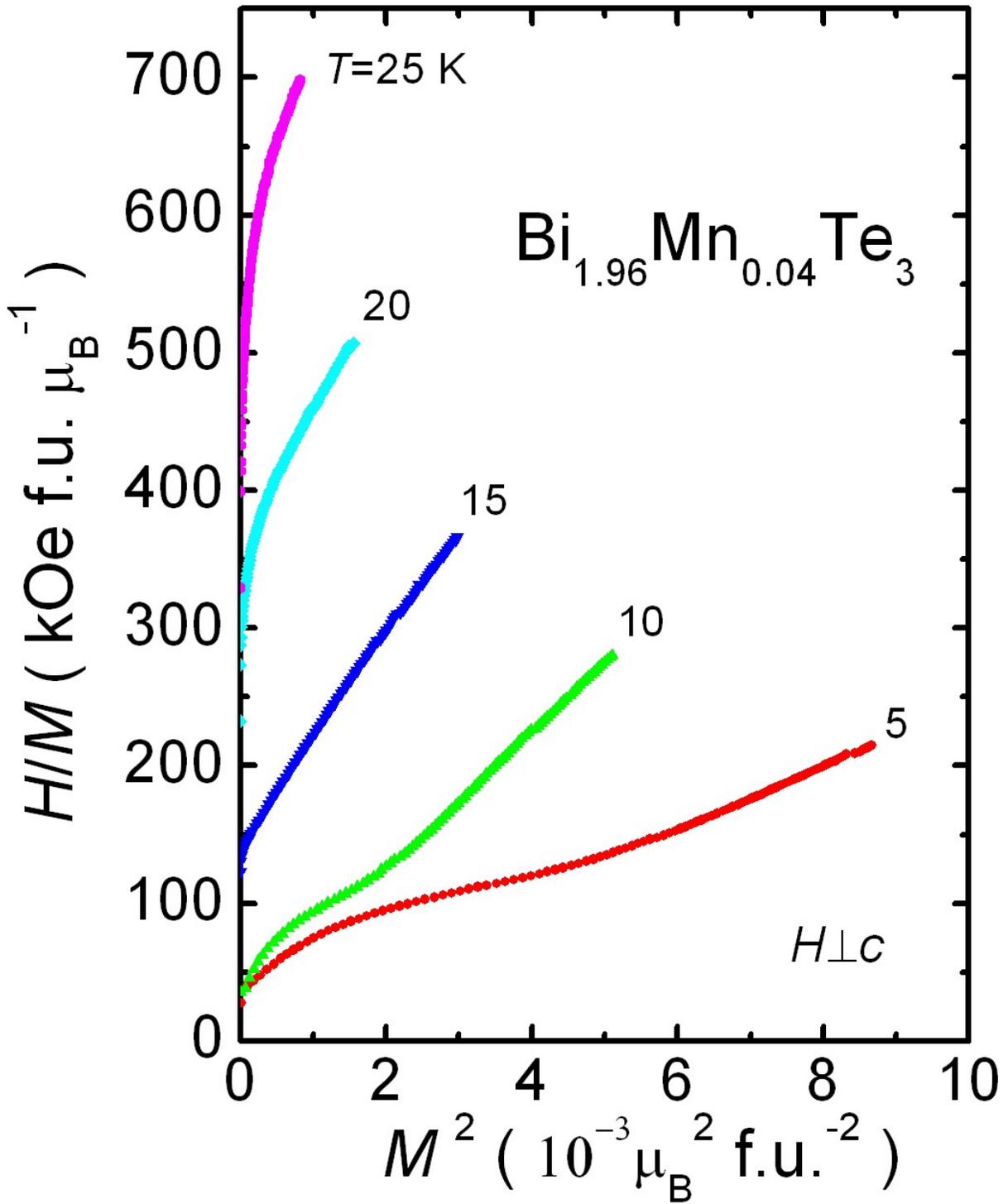

Figure 4

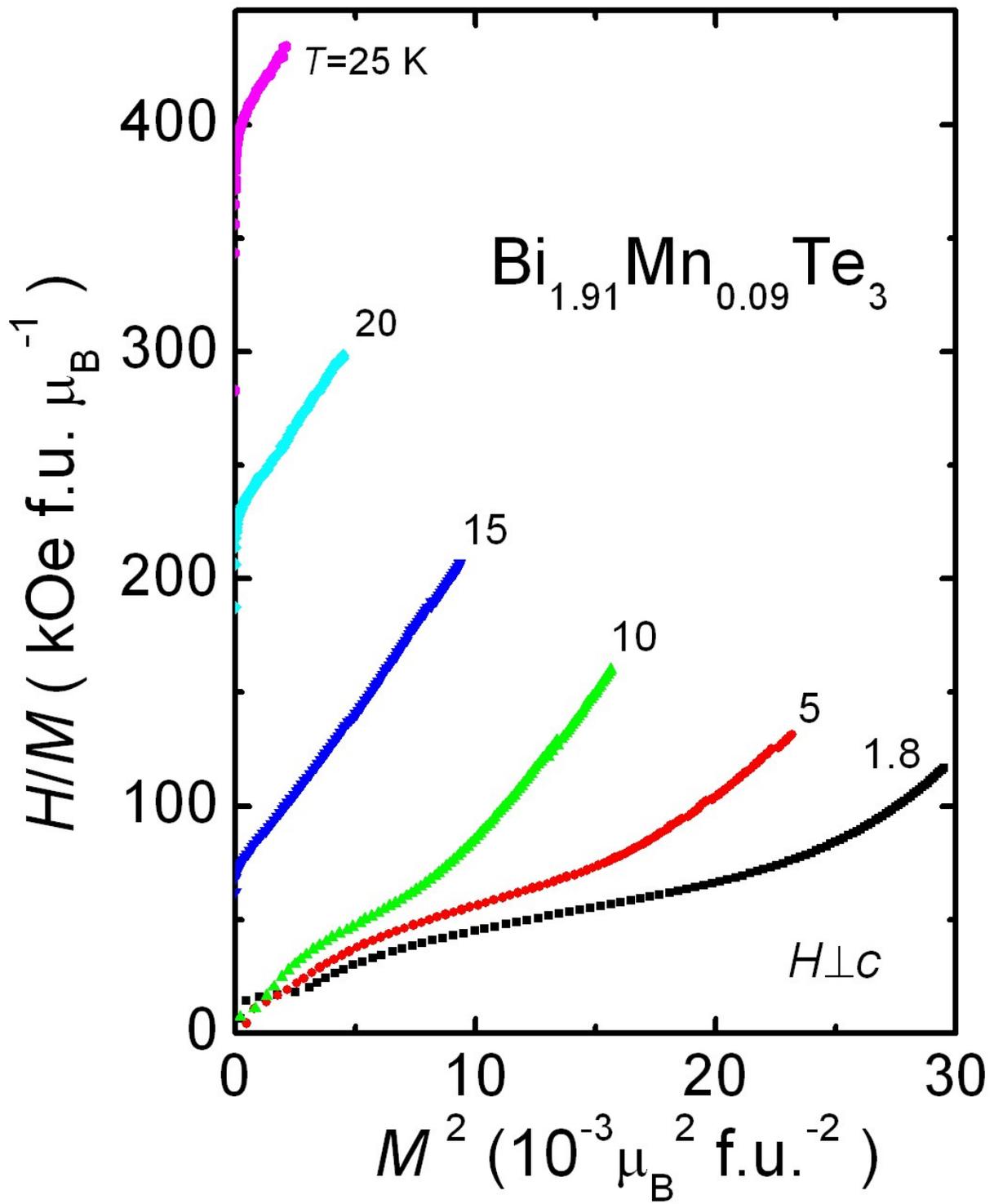

Figure 5

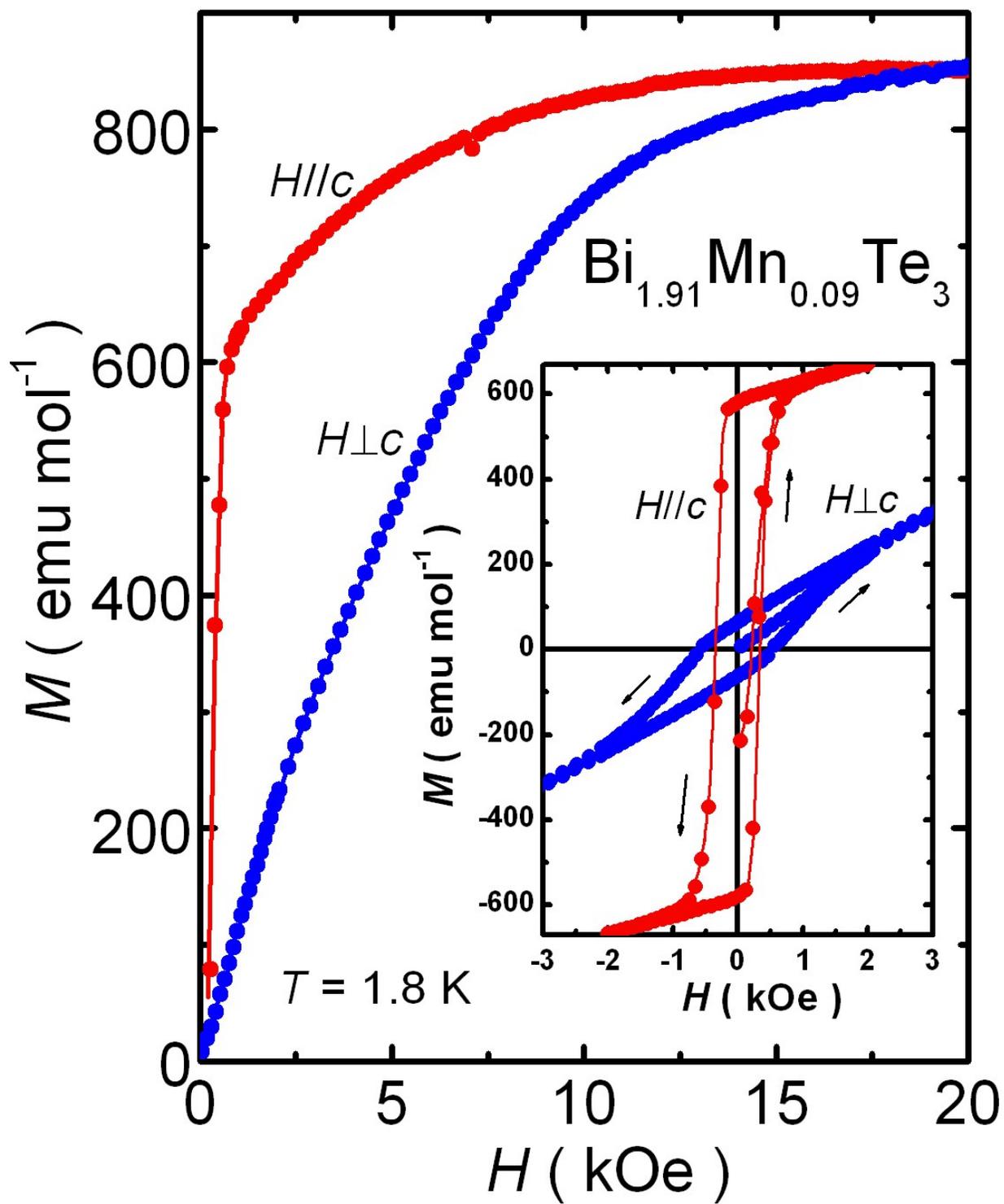

Figure 6

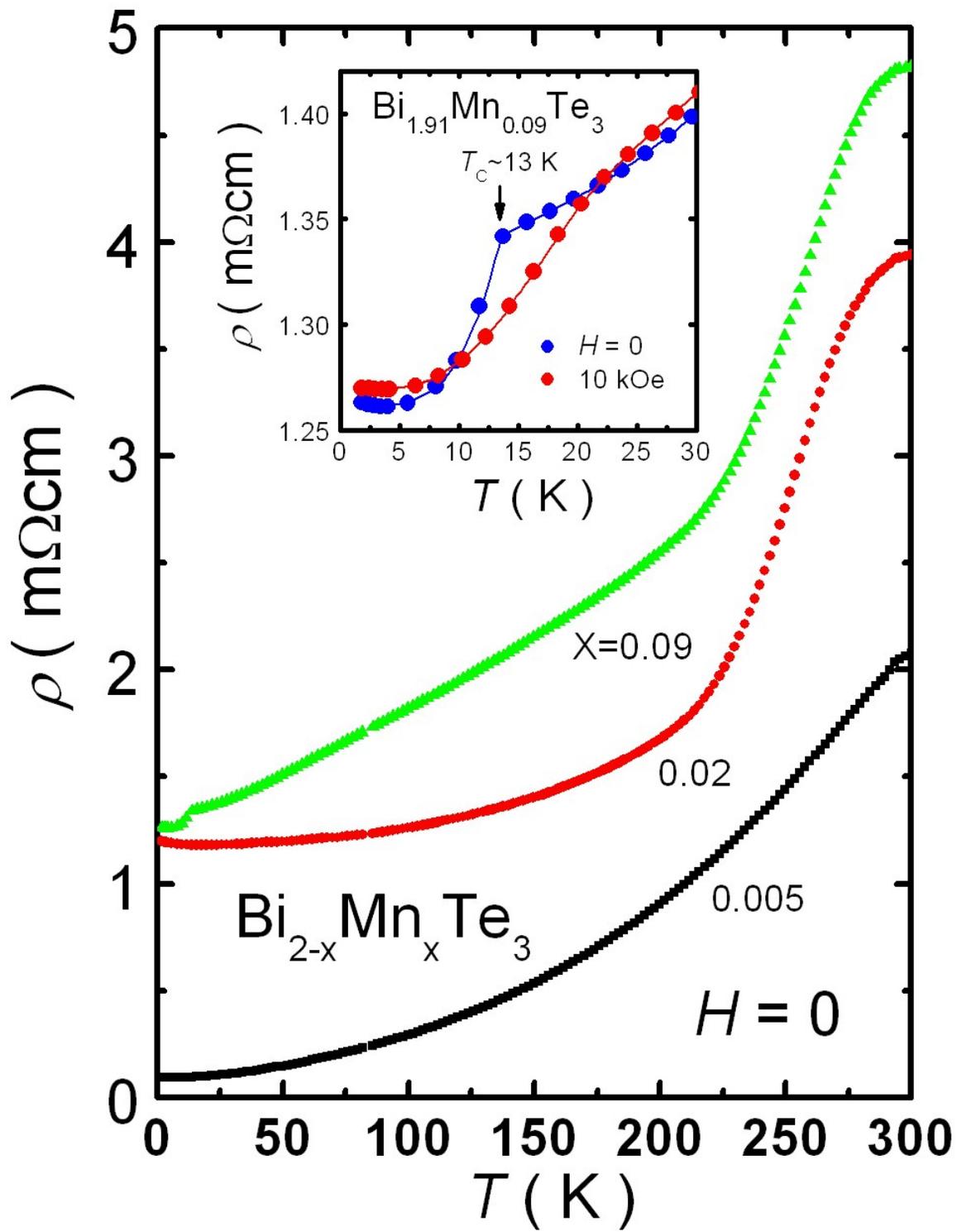

Figure 7

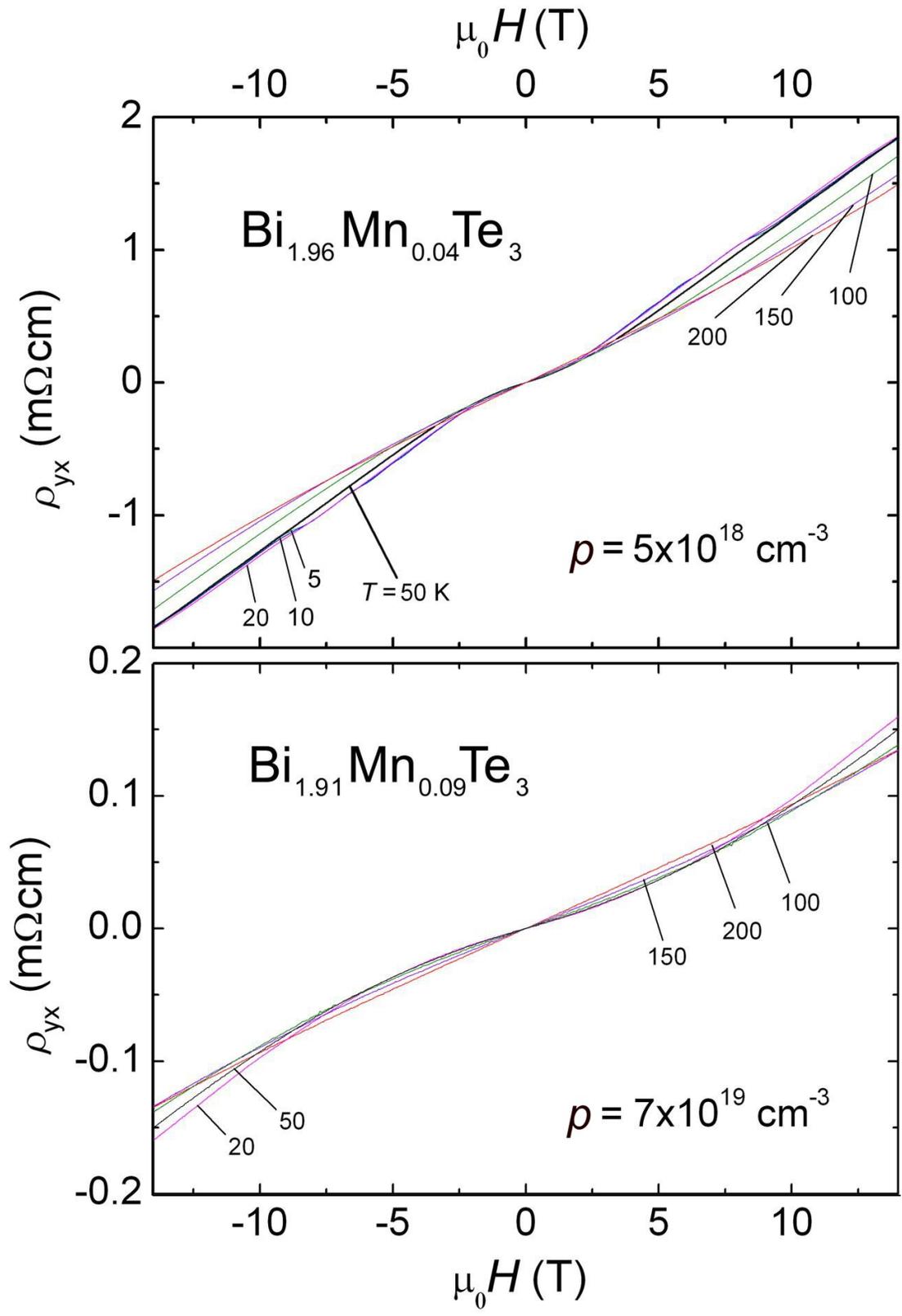

Figure 8

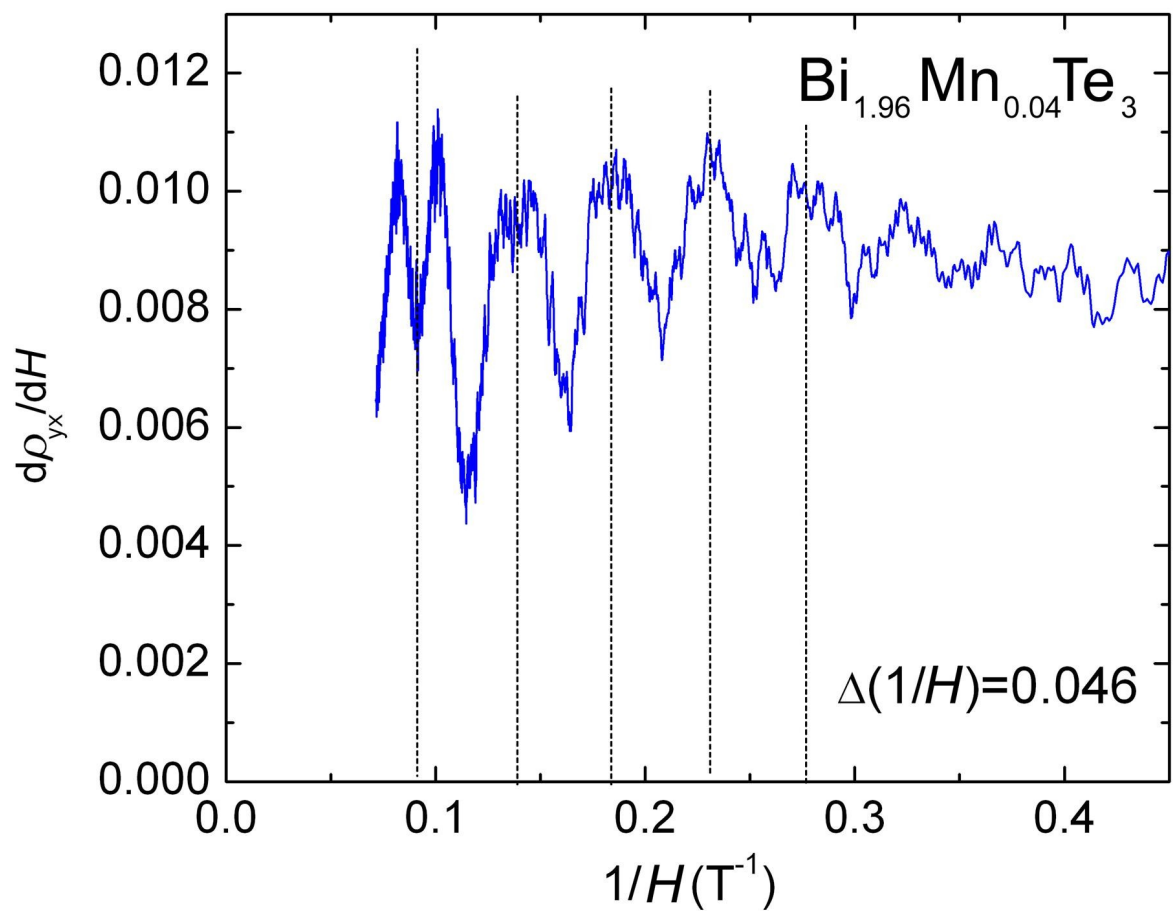

$Bi_{1.96}Mn_{0.04}Te_3$

$\Delta(1/H)=0.046$

$1/H(T^{-1})$

$d\rho_{yx}/dH$

Figure 9

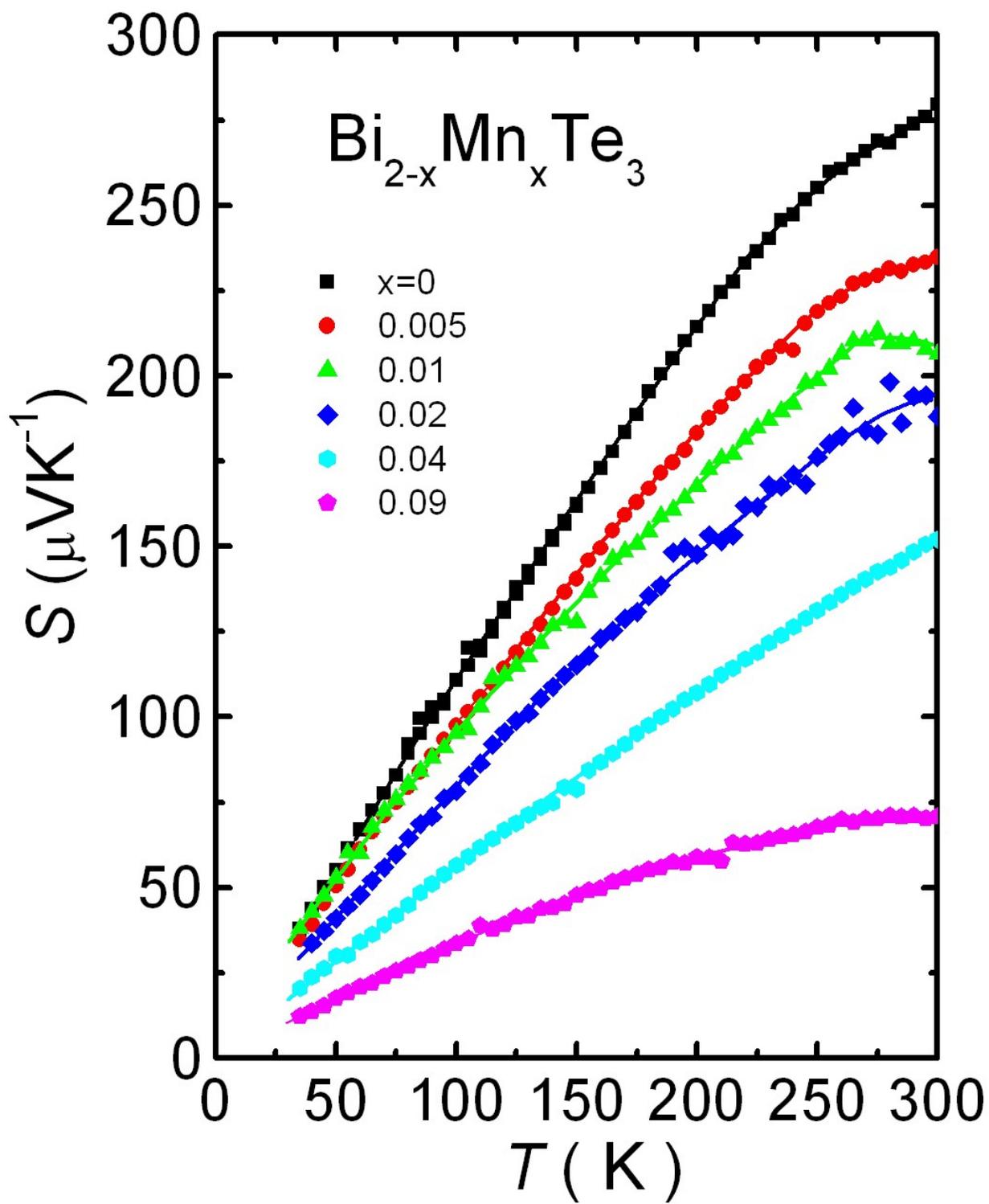

Figure 10

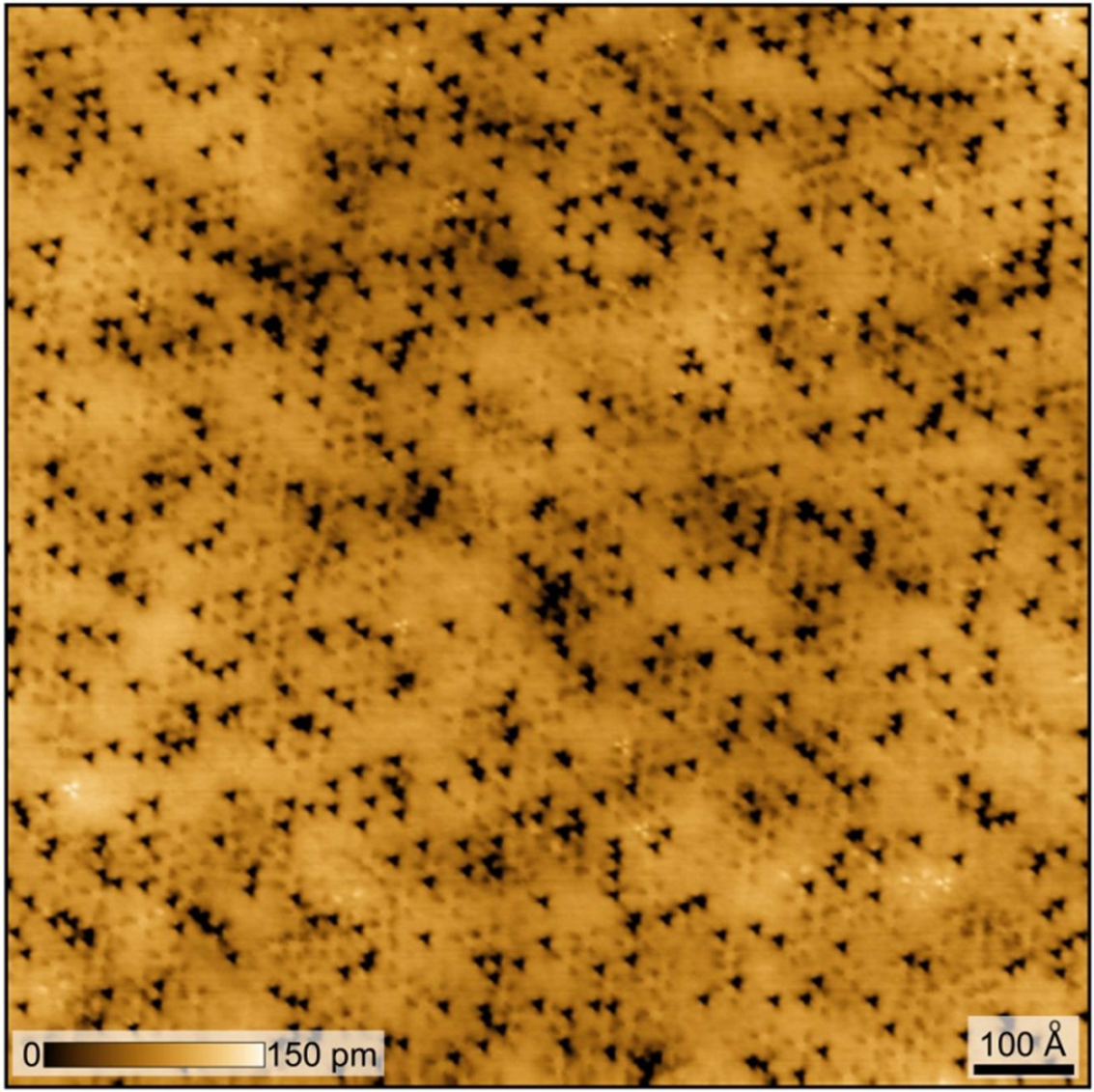

(a)

0      150 pm

100 Å

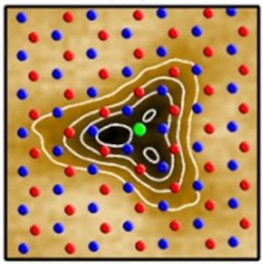

(b) +0.5V

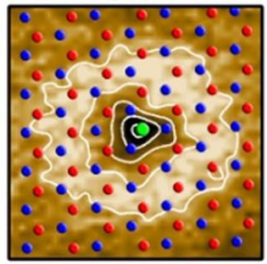

(c) -0.5V

● Te    ● Bi    ● Mn    10 Å

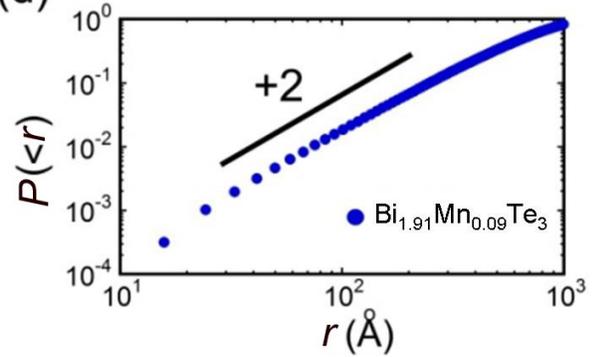

(d)

$P(<r)$

+2

$Bi_{1.91}Mn_{0.09}Te_3$

$r$ (Å)

Figure 11

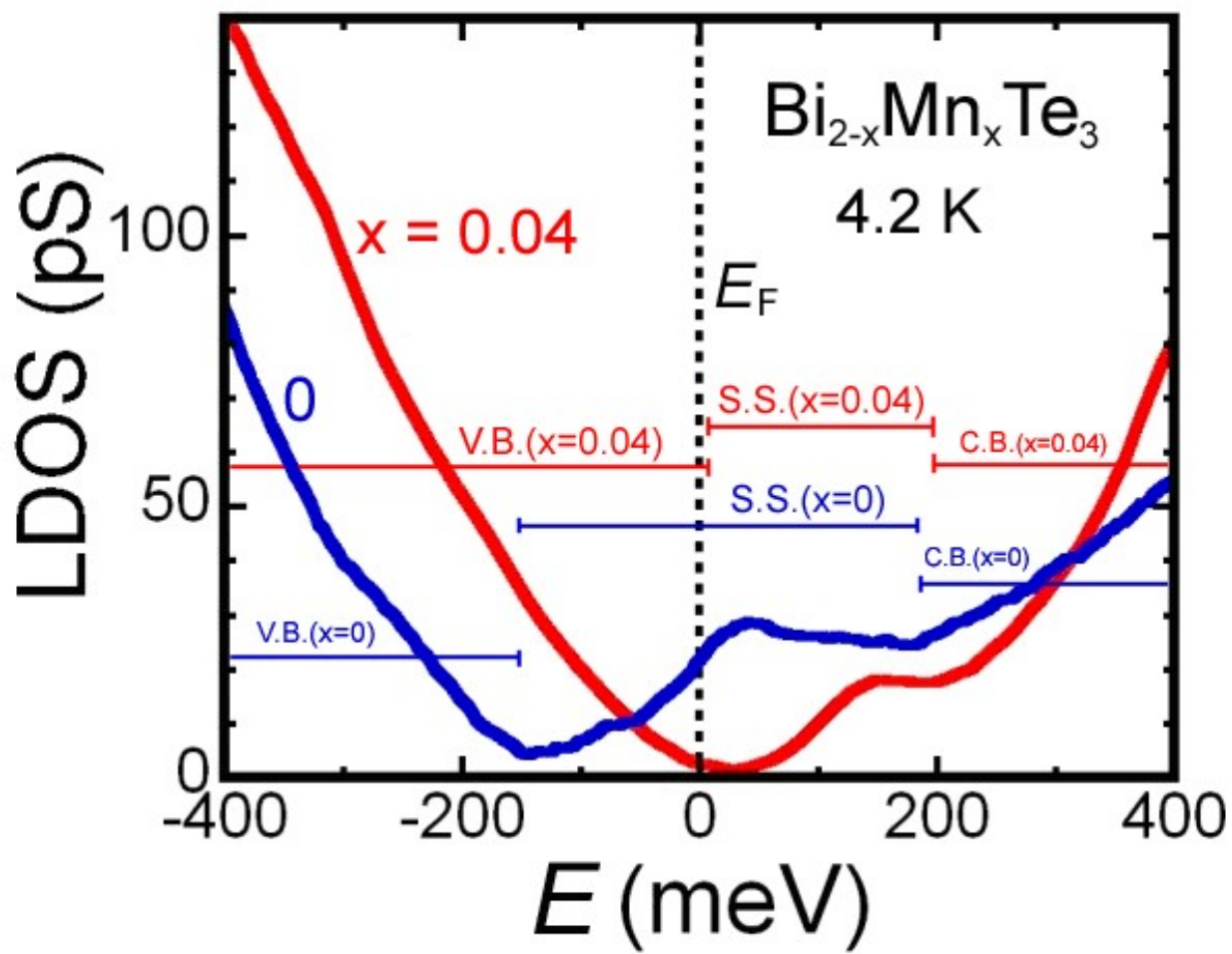

Figure 12

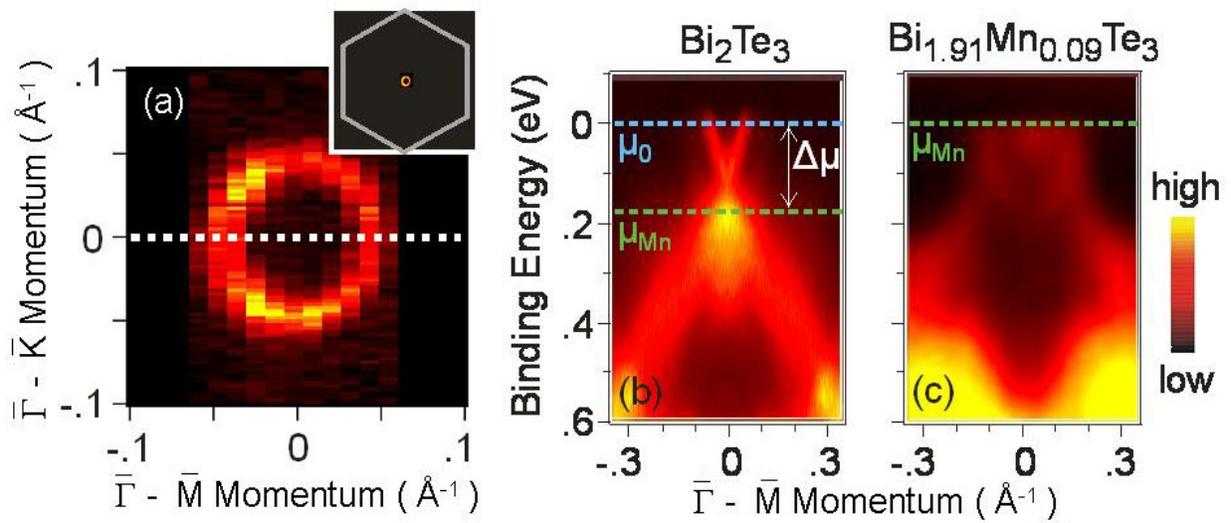

Figure 13